\newcommand{\CLASSINPUTinnersidemargin}{18mm}
\newcommand{\CLASSINPUToutersidemargin}{12mm}
\newcommand{\CLASSINPUTtoptextmargin}{20mm}
\newcommand{\CLASSINPUTbottomtextmargin}{25mm}
\documentclass[10pt,conference,twoside,a4paper]{IEEEtran}
\usepackage[T1]{fontenc}           % Required for T1 fonts
\usepackage{helvet}                % Loads Helvetica (phv)
\usepackage[utf8]{inputenc}        % UTF-8 input
\usepackage{lmodern}               % Optional: better scalable fonts
\usepackage{amsmath,amssymb}       
\usepackage{graphicx}
\usepackage{svg}
\usepackage{subfigure}
\usepackage{booktabs, multirow}
\usepackage{algorithm}
\usepackage{algpseudocode}
\usepackage{xcolor}
\usepackage{url}
\usepackage{hyperref}
\usepackage{amssymb}
\usepackage[utf8]{inputenc}
\usepackage{amsmath} 
\usepackage{hyperref}
\usepackage{algpseudocode}
\usepackage{algorithm}     
\usepackage{times}
\usepackage{abstract}
\renewcommand{\abstractname}{}   
 
\usepackage{graphicx}
\usepackage{subfigure}
\usepackage{svg}
\DeclareGraphicsExtensions{.png,.eps,.ps,.pdf,.svg}
\usepackage{fancyhdr}
\usepackage{kantlipsum}
\usepackage{url}
\usepackage[left=2cm,top=2.5cm,right=2cm,bottom=2.5cm]{geometry}
\usepackage[]{color}
\usepackage{array}
\usepackage{eso-pic}
\begin{document}
\AddToShipoutPictureBG*{%
  \AtPageUpperLeft{%
    \setlength\unitlength{1in}%
 	\hspace{2cm}%% 
 	 	\makebox(0,-2)[l]{%Encabezado página principal
			\begin{tabular}{l r} 
			\multicolumn{1}{p{12cm}}{\vspace{-0.3cm}\includegraphics[scale=0.13]{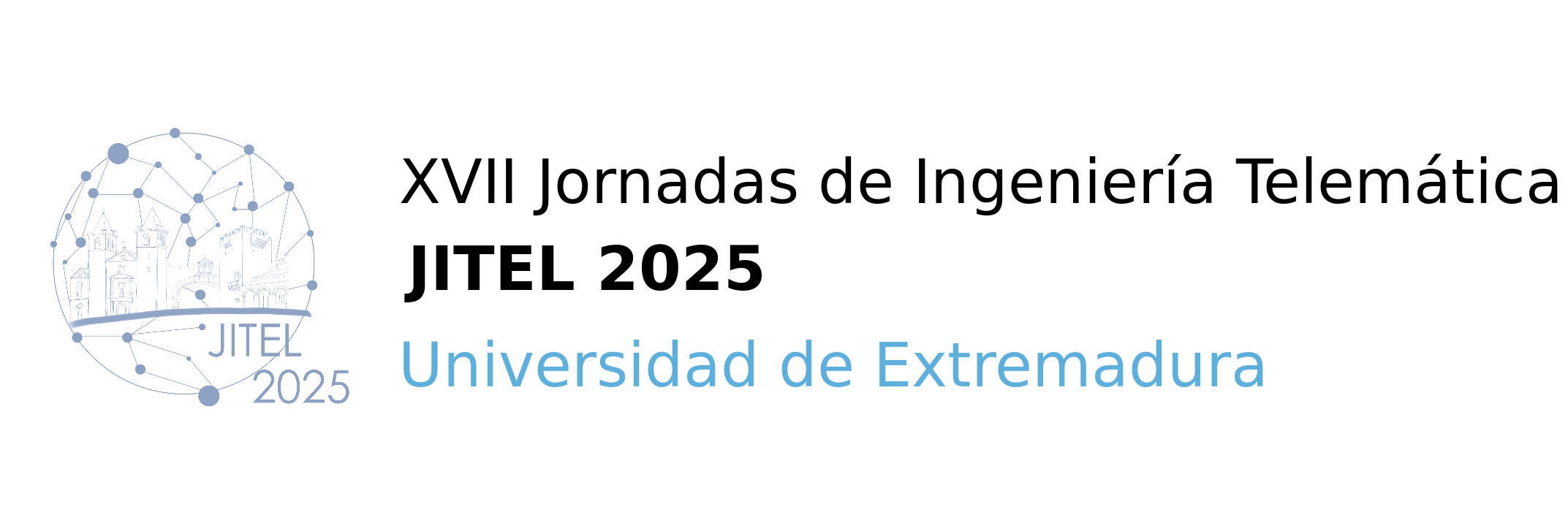}} & \multicolumn{1}{p{4cm}}{\raggedleft\small\usefont{T1}{phv}{m}{it} Actas de las XVII Jornadas de Ingeniería Telemática\\ (JITEL 2025),\\ Cáceres (España), \\12-14 de noviembre de 2025. \vspace{0.2cm} \\ISBN} \tabularnewline 
			\end{tabular}
 }
}}
\AddToShipoutPictureBG*{%
  \AtPageLowerLeft{%
    \setlength\unitlength{1in}%
    \hspace*{\dimexpr0.5\paperwidth\relax}% Pie de página de la página principal.
    \makebox(0,1.3)[c]{\footnotesize\usefont{T1}{phv}{m}{} This work is licensed under a \underline{\textcolor{blue}{Creative Commons 4.0 International License}} (CC BY-NC-ND 4.0)}%
}}

\title{\vspace{3cm}A Reinforcement Learning-Based Telematic Routing Protocol for the Internet of Underwater Things}
\author{
\IEEEauthorblockN{Mohammadhossein Homaei\IEEEauthorrefmark{1}, Mehran Tarif\IEEEauthorrefmark{2}, Agustín Di Bartolo\IEEEauthorrefmark{1},
Víctor González Morales\IEEEauthorrefmark{1},
Mar Ávila Vegas\IEEEauthorrefmark{1}}
\IEEEauthorblockA{\IEEEauthorrefmark{1}Department of Computer Systems Engineering and Telematics, University of Extremadura, \\ Cáceres, 10003, Extremadura, Spain\\
Email: mhomaein@alumnos.unex.es, adibartolo@unex.es, victorgomo@unex.es, 
mmavila@unex.es}
\IEEEauthorblockA{\IEEEauthorrefmark{2}Department of Computer Science, University of Verona, 37134 Verona, Italy\\
Email: mehran.tarifhokmabadi@univr.it}
}

\maketitle

\begin{abstract}
\textbf{
The Internet of Underwater Things (IoUT) has a lot of problems, like low bandwidth, high latency, mobility, and not enough energy.  Routing protocols that were made for land-based networks, like RPL, don't work well in these underwater settings.  This paper talks about RL-RPL-UA, a new routing protocol that uses reinforcement learning to make things work better in underwater situations.  Each node has a small RL agent that picks the best parent node depending on local data such the link quality, buffer level, packet delivery ratio, and remaining energy.  RL-RPL-UA works with all standard RPL messages and adds a dynamic objective function to help people make decisions in real time.  Aqua-Sim simulations demonstrate that RL-RPL-UA boosts packet delivery by up to 9.2\%, uses 14.8\% less energy per packet, and adds 80 seconds to the network's lifetime compared to previous approaches.  These results show that RL-RPL-UA is a potential and energy-efficient way to route data in underwater networks.
}
\end{abstract}


\section{\uppercase{Introduction}}

The IoUT is becoming an important technology for things like monitoring the environment in the ocean, exploring underwater, and checking infrastructure on the ocean floor. These systems use acoustic sensor networks that work in tough conditions with high latency, unstable connections, and severe limits on energy and bandwidth~\cite{Wang2024, Saleem2025}. Underwater networks use acoustic signals instead of radio waves, which are slower and less dependable than radio waves.  Because nodes are generally powered by batteries that are hard to replenish, it's very important that they use as little energy as possible. In addition, node movement can affect the network topology, which makes routing and data delivery uncertain. These problems call for communication methods that are flexible and light.

Many terrestrial IoT devices employ the IPv6 Routing Protocol for Low-Power and Lossy Networks (RPL).  It uses metrics like hop count or Expected Transmission Count (ETX) to create routing trees called Destination-Oriented Directed Acyclic Graphs (DODAGs).  RPL works well in static and low-power situations, but it doesn't work well in dynamic and delay-prone underwater environments since it doesn't change much and doesn't react quickly enough.  In our last paper~\cite{Homaei2025}, we changed RPL's objective function (OF) and communication logic to make it work better underwater.

However, our previous work~\cite{Homaei2025} had some problems: (i) it used static objective function weights that couldn't change when the network conditions changed, (ii) it couldn't learn in real time to make better routing decisions based on past performance, and (iii) it needed to have its parameters manually tuned for each deployment scenario.  These problems show how important it is to find a routing method that is both flexible and energy-efficient, and that works with current RPL systems. We want to use reinforcement learning to build a system like this so that underwater nodes can respond to changing conditions without having to pay a lot for processing or transmission.

 This study builds on our prior work \cite{Homaei2025} to show RL-RPL-UA, a variant of RPL for IoUT that uses RL. A lightweight RL agent runs on each node and chooses the next hop depending on things like energy level, link quality, buffer condition, and delivery history. The protocol uses a composite, adjustable OF to help choose parents while still being fully compatible with RPL control messages (DIO/DAO). RL works well for adaptive routing in networks that change over time because it lets nodes make better judgements based on what they learn.  But most RL-based protocols either don't work with RPL or need a lot of resources. RL-RPL-UA solves this problem by providing a system that can grow and use resources efficiently, and it can adapt to underwater conditions without changing the structure of the RPL protocol.

The rest of the paper is organized as follows: Section~\ref{sec:related} reviews relevant routing protocols for IoUT. Section~\ref{sec:proposed} describes the RL-RPL-UA architecture and objective function. Section~\ref{sec:simulation} presents the simulation setup and performance evaluation. Finally, Section~\ref{sec:conclusion} summarizes conclusions and outlines future work.

 \section{\uppercase{Related Works}}\label{sec:related}

IoUT networks are very hard to utilise because of their specific physical and operational limits, such as high propagation latency, limited energy supplies, frequent disconnections, and dynamic topologies that happen as nodes move around.  There have been several suggestions for routing systems that might fix these issues.  There are five primary groups of these: (i) clustering and depth-based protocols; (ii) game-theoretic and opportunistic strategies; (iii) AI- and RL-driven approaches; (iv) bio-inspired and meta-heuristic algorithms; and (v) extensions based on RPL.

\subsection{Depth-Based and Clustered Routing}

Depth-based and clustered routing approaches try to reduce latency and energy use by putting nodes into clusters or leveraging their depth information. Early protocols like C-GCo-DRAR\cite{Guo2023} and U-(ACH)\textsuperscript{2} \cite{Ismail2023} use node depth and adaptive cluster building to lower latency and transmission overhead. FLCEER employs fuzzy logic~\cite{Natesan2020} to choose the best choice for the cluster head and extend the life of the network.  IDA-OEP employs smart data analytics to transmit data in a way that saves energy ~\cite{Wang2024}, whereas BES uses bald-eagle-search optimisation to make sure that data is sent in the most energy-efficient way ~\cite{Usman2020}. These plans work well in situations that are mobile or very dynamic, but they are typically not adaptable and need to be calibrated to the environment accurately, even though they work well in static conditions.

\subsection{Opportunistic and Game-Theoretic Methods}

Protocols that are opportunistic and game-theoretic deal with void zones, make things more reliable, and consume less energy. For example, GTRP employs Nash equilibria to choose relays in 3-D acoustic networks~\cite{Khan2021}. PCR~\cite{Coutinho2018} changes the power of the gearbox in real time, but hybrid solutions like A-ANTD~\cite{Robinson2020} and TARD~\cite{Saleem2025} use autonomous underwater vehicles (AUVs) to collect data that can handle delays. These designs do make some deployments work better, but they usually need a lot of pre-configuration or centralised control, which stops them from being able to work on their own on a big scale.

\subsection{AI and RL Approaches}

Recent work has employed AI to make routing that adapts to itself.  Li et al. utilise multi-agent RL for optical IoUT links~\cite{Li2020}, Khan et al. use Q-learning for void mitigation~\cite{Khan2021}, and Nandyala et al. create topology-aware Q-routing~\cite{Nandyala2023, Eris2024}.  Tarif et al. use fuzzy inference to stabilise pathways while they are moving.  Tarif et al. (2025) present UWF-RPL, a fuzzy-logic extension of RPL that weights ETX, depth, residual energy, and latency in a Mamdani controller ~\cite{Tarifdm2024,Tariffuzzy2024}. This results in a 17\% PDR gain and 15\% energy savings over baseline RPL~\cite{Tarif2025}.  But unlike the lightweight Q-learning agent we utilised in our RL-RPL-UA, its rule base is fixed and it can't alter weights while it's running.

\subsection{Meta-Heuristic and Bio-Inspired Algorithms}

Bio-inspired and meta-heuristic approaches like FFRP (Firefly)~\cite{Ghoreyshi2018}, EORO (enhanced PSO)~\cite{Ghoreyshi2016}, and BES~\cite{Usman2020} employ swarm intelligence to find paths that consume less energy. Pradeep et al. \cite{Pradeep2023} provide a fuzzy region-based approach that works with sink mobility. Even if their simulation results are promising, they often can't be used in the real world since they need global optimisation and don't learn continuously.

\subsection{RPL-Based Extensions}

The RPL was first created for sensor networks on land, but it has now been changed to function underwater.  Because RPL is the primary protocol for IoT on land, a lot of research has gone into making it work with underwater acoustic communication.  In UW/MRPL~\cite{Homaei2025}, we made RPL work better in underwater settings by adding routing metrics that take depth into account and support for mobility.  It was better than basic protocols like OF0 and MRHOF, but it couldn't adjust in real time and used fixed objective function (OF) weights.  To solve the problems of compatibility and energy balancing, UWF-RPL~\cite{Tarif2025} added a fuzzy logic-based OF to conventional RPL control messages (DIO/DAO).  It didn't, meanwhile, use feedback systems to make routing decisions better, and membership functions still had to be changed by hand.
 Our proposed RL-RPL-UA improves RPL by adding an RL agent that automatically changes OF weights in real time. This keeps complete protocol compatibility and gets rid of the requirement for human configuration.

\begin{table}[ht!]
\centering
\caption{Comparison of Recent Routing Protocols with RL-RPL-UA}
\label{tab:comparison1}
\resizebox{\columnwidth}{!}{%
\begin{tabular}{l c c c c c}
\hline
\hline
\textbf{Protocol} & \textbf{RL} & \textbf{RPL} & \textbf{Adaptive OF} & \textbf{Mobility} & \textbf{Citation} \\
\hline
C-GCo-DRAR & -- & -- & Static OF & Limited & \cite{Guo2023} \\
FLCEER & -- & -- & Static OF & Moderate & \cite{Natesan2020} \\
IDA-OEP & -- & -- & Static OF & Limited & \cite{Wang2024} \\
GTRP & -- & -- & Static OF & Moderate & \cite{Khan2021} \\

RL Protocol & \checkmark & -- &  Static OF & Moderate &  \cite{Eris2024}\\
Q-Learning & \checkmark & -- & Dynamic OF & Moderate & \cite{Nandyala2023} \\
Multi-agent RL & \checkmark & -- & Static OF & High & \cite{Li2020} \\
UA-RPL & -- & \checkmark & Static OF & Moderate & \cite{Liu2022} \\
URPL & -- & \checkmark & Dynamic OF & Moderate & \cite{Tarifdm2024} \\
Fuzzy-CR & -- & \checkmark & Decision Making & Moderate & \cite{Tariffuzzy2024} \\
UWF-RPL & -- & \checkmark & Static Fuzzy OF & Moderate & \cite{Tarif2025} \\

UW/MRPL (prev. work) & -- & \checkmark & Static OF & High & \cite{Homaei2025} \\
RL-RPL-UA (this work) & \checkmark & \checkmark & Dynamic & High & -- \\
\hline
\hline
\end{tabular}%
}
\end{table}

\vspace{0.5em}
We briefly review them above, and use Tables~\ref{tab:comparison1} and~\ref{tab:comparison2} to illustrate why RL-RPL-UA is necessary and novel.

\begin{table*}[t!]
\scriptsize
\renewcommand{\arraystretch}{1.2}
\centering
\caption{Key Differences Between UWF-RPL, UWMRPL (Our previous works), and RL-RPL-UA}
\label{tab:comparison2}
\begin{tabular}{>{\centering\arraybackslash}p{2.5cm} >{\centering\arraybackslash}p{4.5cm} >{\centering\arraybackslash}p{4.5cm} >{\centering\arraybackslash}p{4.5cm}}
\hline
\hline
\textbf{Feature} & \textbf{UWF-RPL \cite{Tarif2025}} & \textbf{UWMRPL \cite{Homaei2025} (Previous work)} & \textbf{RL-RPL-UA (This Work)} \\
\hline
Main Concept & Fuzzy-logic RPL for optimized routing & Mobility-aware RPL with static tunable OF & RL-based dynamic routing with local agents \\
Routing Adaptability & Semi-adaptive via static fuzzy logic rules & Semi-adaptive via predefined logic & Fully adaptive via real-time RL updates \\
OF & Static fuzzy logic (depth, energy, latency, ETX) & Static/custom (ETX, depth) & Dynamic composite (energy, LQI, queue, PDR) \\
RPL Compatibility & Extended DIO/DAO with fuzzy logic metrics & Standard-compliant & Extended DIO/DAO with RL metrics \\
Learning Agent & None (static fuzzy logic) & None & Q-learning or DQN per node \\
Reward Mechanism & None & None & $\alpha \cdot \text{PDR} - \beta \cdot \text{Delay} - \gamma \cdot \text{Cost}$ \\
Overhead & Moderate (fuzzy logic computations) & Low (no learning updates) & Low (optimized RL updates) \\
Mobility Handling & Reactive via fuzzy logic evaluation & Reactive DAG repair & Proactive via learned feedback \\
Queue Management & Included (congestion control) & Not included & Included (adaptive queue management) \\
Energy Efficiency & Good (static optimized) & Moderate (no dynamic optimization) & High (real-time optimization) \\
Key Contribution & Improved stability and efficiency via fuzzy logic & Mobility and energy-aware extension of RPL & Online adaptive parent selection via RL \\
\hline
\hline
\end{tabular}
\end{table*}

\section{\uppercase{Proposed Protocol: RL-RPL-UA}}\label{sec:proposed}

In this section, we introduce RL-RPL-UA, a novel routing protocol that enhances the conventional RPL protocol by embedding an RL agent into each node of the underwater IoT network. Unlike traditional RPL implementations that rely on static OFs, our model leverages an adaptive learning mechanism to select optimal routes under the harsh and dynamic conditions of underwater communication.

\subsection{Protocol Architecture}
The architecture of RL-RPL-UA integrates an RL agent within the standard RPL stack. Each node consists of the following modules:

\begin{itemize}
    \item \textbf{Sensing Unit:} Gathers state information including residual energy, buffer occupancy, and signal strength.
    \item \textbf{Communication Module:} Interfaces with an acoustic modem or underwater simulation module (Aqua-Sim, NS-2 with UAN).
    \item \textbf{RL Agent:} A local Q-learning or DQN model.
    \item \textbf{Extended RPL Stack:} Supports modified DIO/DAO messages carrying dynamic state and learned metrics.
\end{itemize}

\subsection{RL Model}

The routing process is modeled as a Markov Decision Process (MDP), where each node learns an optimal routing policy by interacting with its environment.

\subsubsection{State Space}

The state $s_t$ at time $t$ includes:

\begin{equation}
s_t = \left[E_t, \text{LQI}_t, Q_t, \text{PDR}_t, T_t\right]
\label{eq:state}
\end{equation}

where $E_t$ is residual energy, $\text{LQI}_t$ is link quality indicator, $Q_t$ is current queue size, $\text{PDR}_t$ is historical packet delivery ratio, and $T_t$ is time since last successful transmission.

\subsubsection{Action Space}

The action $a_t$ is the selection of a next-hop parent from among $n$ neighbors:

\begin{equation}
a_t \in \{ \text{Parent}_1, \text{Parent}_2, \dots, \text{Parent}_n \}
\label{eq:action}
\end{equation}

As shown in Equation~\eqref{eq:action}, the action space consists of the set of all neighboring nodes that can serve as the next hop in the routing process.

\subsubsection{Reward Function}

The reward signal $r_t$ is defined to balance reliability, delay, and energy consumption:

\begin{equation}
r_t = \alpha \cdot \text{PDR}_t - \beta \cdot \text{Delay}_t - \gamma \cdot \text{EnergyCost}_t
\label{eq:reward}
\end{equation}

As shown in Equation~\eqref{eq:reward}, this formulation enables the agent to optimize routing decisions by weighing the positive effect of packet delivery against the negative impact of delay and energy consumption. Here, $\alpha$, $\beta$, and $\gamma$ are tunable hyperparameters that control the trade-offs between these objectives. This reward is used to update the RL agent's policy.

\begin{algorithm}[H]
\scriptsize
\caption{RL-enhanced RPL Routing}
\label{alg:rl-rpl-ua}
\begin{algorithmic}[1]

\Require Initialization of Q-table or DQN weights, neighbor table, default Rank
\Ensure Energy-efficient and adaptive routing in underwater IoT

\State \textbf{Initialize} RL agent (Q-table or DQN), default Rank
\State $s \gets \Call{observe\_state}{}$
\State \textbf{Broadcast} DIO with $OF_{RL}(n_i)$ and node state

\While{Node is active}

    \State \textbf{Receive DIOs} from neighbors

    \ForAll{neighbor $n_i$ in NeighborTable}
        \State Extract state features: $s_{n_i} = [E, LQI, Q, PDR, T]$
        \State Compute $OF_{RL}(n_i)$ using Equation~\ref{eq:objective}
        \State Estimate $Q(s, a=n_i)$ using RL model (Q-table or DQN)
    \EndFor

    \State \textbf{Select Parent}:
    \State $a^* \gets \arg\max_{n_i} Q(s, a=n_i)$ \Comment{Best next-hop based on RL}

    \State \textbf{Update RPL Rank} based on selected parent and $OF_{RL}(a^*)$
    \State \textbf{Forward data} packets to $a^*$

    \State Wait for Acknowledgement or Timeout

    \State \textbf{Observe outcome:}
    \State Measure $\text{PDR}_t$, $\text{Delay}_t$, $\text{EnergyCost}_t$
    \State Compute reward $r_t$ using Equation~\ref{eq:reward}
    \State $s_{t+1} \gets \Call{observe\_state}{}$

    \State \textbf{RL Update:}
    \If{Using Q-learning}
        \State Update Q-table using Equation~\ref{eq:qlearning}
    \ElsIf{Using DQN}
        \State Store $(s_t, a^*, r_t, s_{t+1})$ in ReplayBuffer
        \State Train DQN via minibatch sampling
    \EndIf

    \State $s_t \gets s_{t+1}$
    \State \textbf{Periodically broadcast updated DIO} with new Rank and $OF_{RL}$

\EndWhile

\Function{observe\_state}{}
    \State Measure local energy $E$, link quality $LQI$, buffer queue $Q$, packet success rate $PDR$, time since last ACK $T$
    \State \Return $[E, LQI, Q, PDR, T]$
\EndFunction

\end{algorithmic}
\end{algorithm}

\subsubsection{Policy Learning}

The RL agent seeks to learn a policy $\pi(a|s)$ that maximizes the expected cumulative reward:
\begin{align}
Q(s_t, a_t) \leftarrow\; & Q(s_t, a_t) + \eta \Big[ r_t + \gamma \cdot \max_{a'} Q(s_{t+1}, a') \notag \\
& - Q(s_t, a_t) \Big]
\label{eq:qlearning}
\end{align}

Equation \ref{eq:qlearning} is the standard Q-learning update rule, where $\eta$ is the learning rate and $\gamma$ is the discount factor for future rewards.

\subsection{Adaptive OF}
To replace the static OFs in RPL, we define a composite and dynamic OF:
\begin{align}
OF_{RL}(n_i) =\; & w_1 \cdot E(n_i) + w_2 \cdot R(n_i) \notag \\
& + w_3 \cdot Q(n_i) + w_4 \cdot PDR(n_i)
\label{eq:objective}
\end{align}

Where:
\begin{itemize}
    \item $E(n_i)$: Normalized remaining energy of neighbor $n_i$,
    \item $R(n_i)$: Link reliability (e.g., inverse of ETX),
    \item $Q(n_i)$: Queue length or buffer utilization,
    \item $PDR(n_i)$: Historical delivery ratio,
    \item $w_1$ to $w_4$: Adaptive weights tuned by the RL agent.
\end{itemize}

This OF is broadcast in DIO messages, allowing each node to evaluate its neighbors and update its rank dynamically.

\subsection{Routing Decision Process}

The routing process in RL-RPL-UA involves the following steps:

\begin{enumerate}
    \item DIO Exchange: Each node broadcasts its current state and $OF_{RL}$ value using an extended DIO message.
    \item Neighbor Table Update: On receiving DIOs, nodes update their neighbor tables and estimate Q-values.
    \item Parent Selection: The parent with the highest Q-value is selected as the preferred next-hop.
    \item Data Forwarding: Data packets are forwarded along the learned optimal path.
    \item Learning Update: After each transmission, the node observes outcomes and updates its Q-table using Equation \ref{eq:qlearning}.
\end{enumerate}

\subsection{Underwater-Specific Enhancements}

The following improvements are tailored to the underwater environment:

\begin{itemize}
    \item Delay Estimation: Nodes estimate propagation delay based on distance and water temperature to better model the reward function.
    \item Energy-Aware Slot Scheduling: TDMA (Time Division Multiple Access) is used as a MAC protocol to assign non-overlapping time slots to nodes, reducing collisions and idle listening in underwater acoustic networks.
    \item Clustered Learning: In large networks, cluster heads can aggregate policies and periodically disseminate updates.
\end{itemize}

\subsection{Compatibility and Overhead}
RL-RPL-UA remains compatible with legacy RPL nodes by embedding new fields in the optional sections of RPL messages. In terms of complexity:
\begin{itemize}
    \item The Q-learning implementation requires minimal computational resources and is suitable for constrained devices.
    \item Communication overhead is slightly increased due to additional state sharing, but overall packet retransmissions are reduced.
\end{itemize}

\subsection{Security Considerations}
RL-RPL-UA can be extended to support trust-aware routing by integrating reputation scores into the reward function, allowing the network to avoid compromised nodes.

\subsection{Resource and Energy Cost Estimation}

To evaluate the feasibility of deploying RL-RPL-UA in real-world IoUT scenarios, we estimate the energy and processing cost based on standard underwater sensor node specifications. We consider nodes equipped with low-power microcontrollers (e.g., MSP430, ARM Cortex-M4) and acoustic modems such as the WHOI Micromodem or EvoLogics S2C.

\subsubsection{Energy Cost per Transmission}
Assuming a transmission power of 0.5 W and transmission time of 1.5 seconds per packet, the energy cost per transmission is calculated as:

\begin{equation}
E_{tx} = P_{tx} \times t = 0.5 \times 1.5 = 0.75 \text{ J}
\label{eq:etx}
\end{equation}

As shown in Equation~\eqref{eq:etx}, each data transmission consumes approximately 0.75 joules.

\subsubsection{Energy Cost per RL Update}

The Q-table update process requires approximately 500–1000 CPU cycles. For a 16 MHz processor operating at 1.8 V and 3 mA, the energy cost is given by:

\begin{equation}
E_{cpu} = V \times I \times \frac{\text{cycles}}{f} = 1.8 \times 0.003 \times \frac{1000}{16 \times 10^6} \approx 0.34 \mu\text{J}
\label{eq:ecpu}
\end{equation}

According to Equation~\eqref{eq:ecpu}, the energy consumption for a single RL update is approximately 0.34 microjoules.
\subsubsection{Memory and Storage}

The Q-table of size $n \times a$ with 8-bit values, for example with 10 neighbors and 5 actions, requires approximately $50$ bytes. This is feasible for microcontrollers with at least $32$~KB of SRAM.

\subsubsection{Discussion}

Compared to traditional RPL, RL-RPL-UA introduces minimal computational overhead due to the small Q-table and low RL update cost. However, it improves energy efficiency by reducing retransmissions and adapting paths proactively.

\section{\uppercase{Simulation Results}}\label{sec:simulation}

\subsection{\uppercase{Simulation Parameters}}
To assess the performance of the proposed RL-RPL-UA protocol, we conducted simulations using Aqua-Sim, an extension of the NS-2 (Network Simulator 2) framework specifically designed for underwater acoustic network environments \cite{Xie2009}. Aqua-Sim is available at \url{https://github.com/rmartin5/aqua-sim-ng}. The simulated network consists of both static and mobile sensor nodes deployed within a 3D underwater space using acoustic communication links. Each node independently executes the RL-RPL-UA algorithm and exchanges routing information via modified DIO/DAO messages.

We compare RL-RPL-UA against several baseline protocols, including standard RPL (OF0), Q-learning-only approaches, and cluster-based routing. We specifically selected Co-DRAR~\cite{Guo2023} as it represents state-of-the-art depth-based clustered routing, UA-RPL~\cite{Liu2022} as a recent RPL adaptation for underwater networks, and UWF-RPL~\cite{Tarif2025} as the most recent fuzzy-logic enhancement of RPL, along with our prior work UW/MRPL~\cite{Homaei2025}. These protocols were chosen because they cover different routing paradigms (clustered, RPL-based, fuzzy-enhanced, and mobility-aware) and allow comprehensive evaluation of RL-RPL-UA's improvements in adaptability, energy efficiency, and delivery reliability. Evaluation metrics include Packet Delivery Ratio (PDR), End-to-End Delay, Energy Consumption, Routing Overhead, and Network Lifetime.

The main simulation parameters are listed in Table~\ref{tab:simparams}.

\begin{table}[ht]
\centering
\caption{Simulation Parameters for RL-RPL-UA Evaluation}
\label{tab:simparams}
\begin{tabular}{l c}
\hline
\hline
\textbf{Parameter} & \textbf{Value} \\
\hline
Simulation Area & $300 \times 300 \times 300$ m$^3$ \\
Number of Nodes & 50 \\
Initial Energy per Node & 5 J \\
Transmission Power & 0.5 W \\
Acoustic Bandwidth & 10 kHz \\
Propagation Speed & 1500 m/s \\
MAC Protocol & TDMA \\
Routing Protocols & RL-RPL-UA, RPL (OF0), Q-Routing \\
RL Algorithm & Q-learning (tabular) \\
Learning Rate $(\eta)$ & 0.1 \\
Discount Factor $(\gamma)$ & 0.9 \\
Reward Weights $(\alpha, \beta, \gamma)$ & (1.0, 0.6, 0.4) \\
Simulation Time & 1000 s \\
Traffic Model & CBR, 1 packet/10 s \\
Packet Size & 64 Bytes \\
Mobility Model & Random Waypoint (0.1–0.3 m/s) \\
\hline
\hline
\end{tabular}
\end{table}

\subsection{Packet Delivery Ratio}

PDR is calculated over $K$ simulation trials as \cite{Homaei2021}:

\begin{equation}
\text{PDR}_{\text{mean}} = \frac{1}{K} \sum_{k=1}^{K} 
\left( \frac{R_k}{S_k} \right) \times 100
\end{equation}

\begin{equation}
\sigma_{\text{PDR}} = \sqrt{ \frac{1}{K-1} \sum_{k=1}^{K} 
\left( \frac{R_k}{S_k} - \text{PDR}_{\text{mean}} \right)^2 }
\end{equation}

In the static scenario, RL-RPL-UA achieves a mean Packet Delivery Ratio (PDR) of 94.3\% with a standard deviation of 1.7, outperforming UWF-RPL (89.2\%, $\sigma$=2.2), UWRPL (85.1\%, $\sigma$=3.0), UA-RPL (83.5\%, $\sigma$=3.0), and Co-DRAR (81.2\%, $\sigma$=2.8). The results show that while UWF-RPL enhances PDR over traditional RPL variants by using adaptive metrics, RL-RPL-UA delivers a further 5.1\% improvement over UWF-RPL and 9.2\% over UWRPL, confirming the impact of RL in static deployments.

In the mobile scenario, RL-RPL-UA maintains the highest delivery performance with a PDR of 92.8\% ($\sigma$=1.9), surpassing UWF-RPL (90.5\%, $\sigma$=2.0), UWMRPL (88.2\%, $\sigma$=2.3), UA-RPL (80.2\%, $\sigma$=3.1), and Co-DRAR (78.4\%, $\sigma$=3.2). Although UWF-RPL narrows the performance gap in mobile conditions through fuzzy logic and energy-aware decisions, RL-RPL-UA outperforms all baselines, confirming that its real-time learning approach significantly improves delivery reliability under dynamic underwater environments.

\begin{figure}[ht]
    \centering
    \includegraphics[width=1\linewidth]{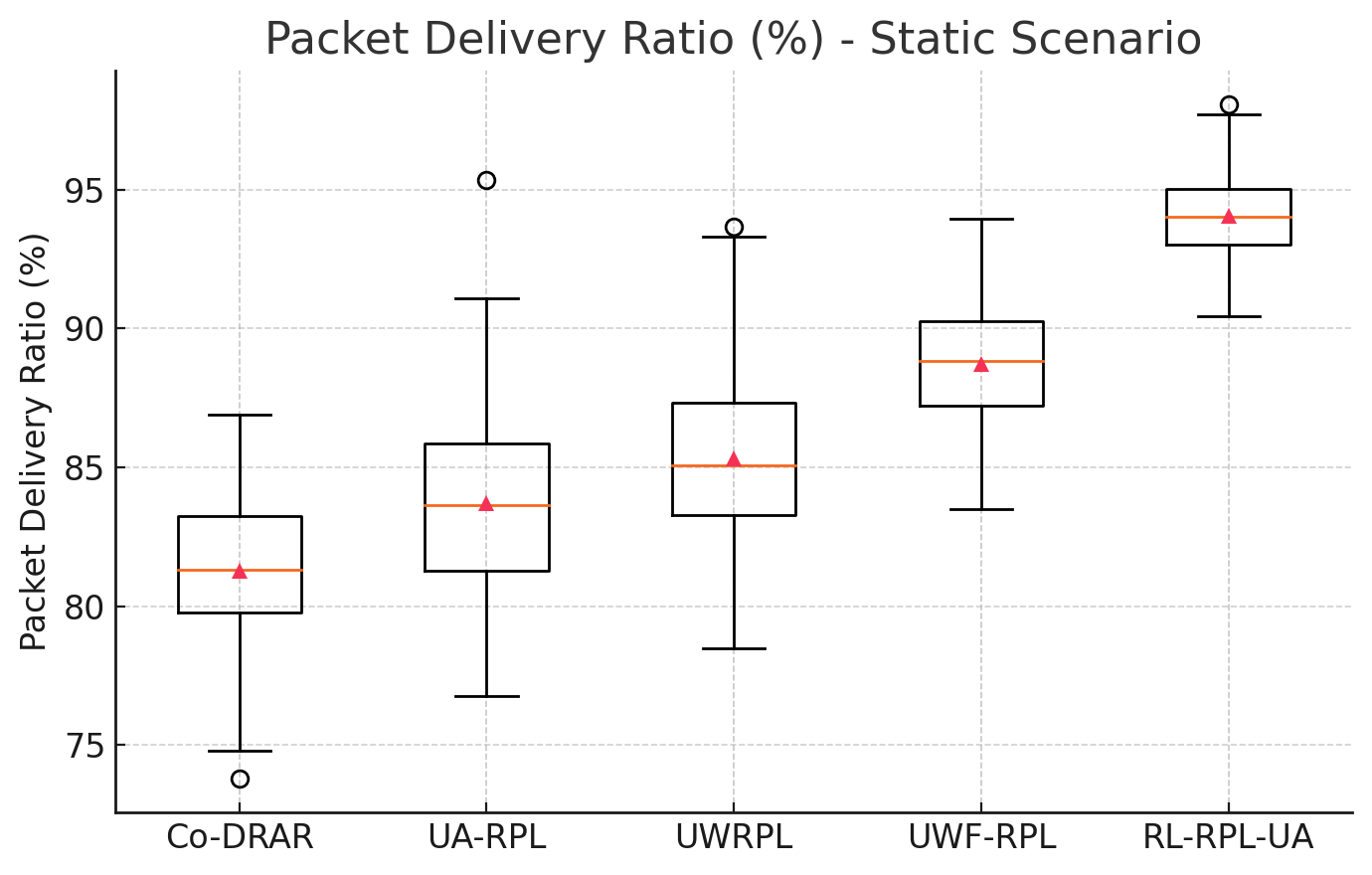}
    \caption{PDR in the static scenario.}
    \label{fig:PDR_Static}
\end{figure}

\begin{figure}[ht]
    \centering
    \includegraphics[width=1\linewidth]{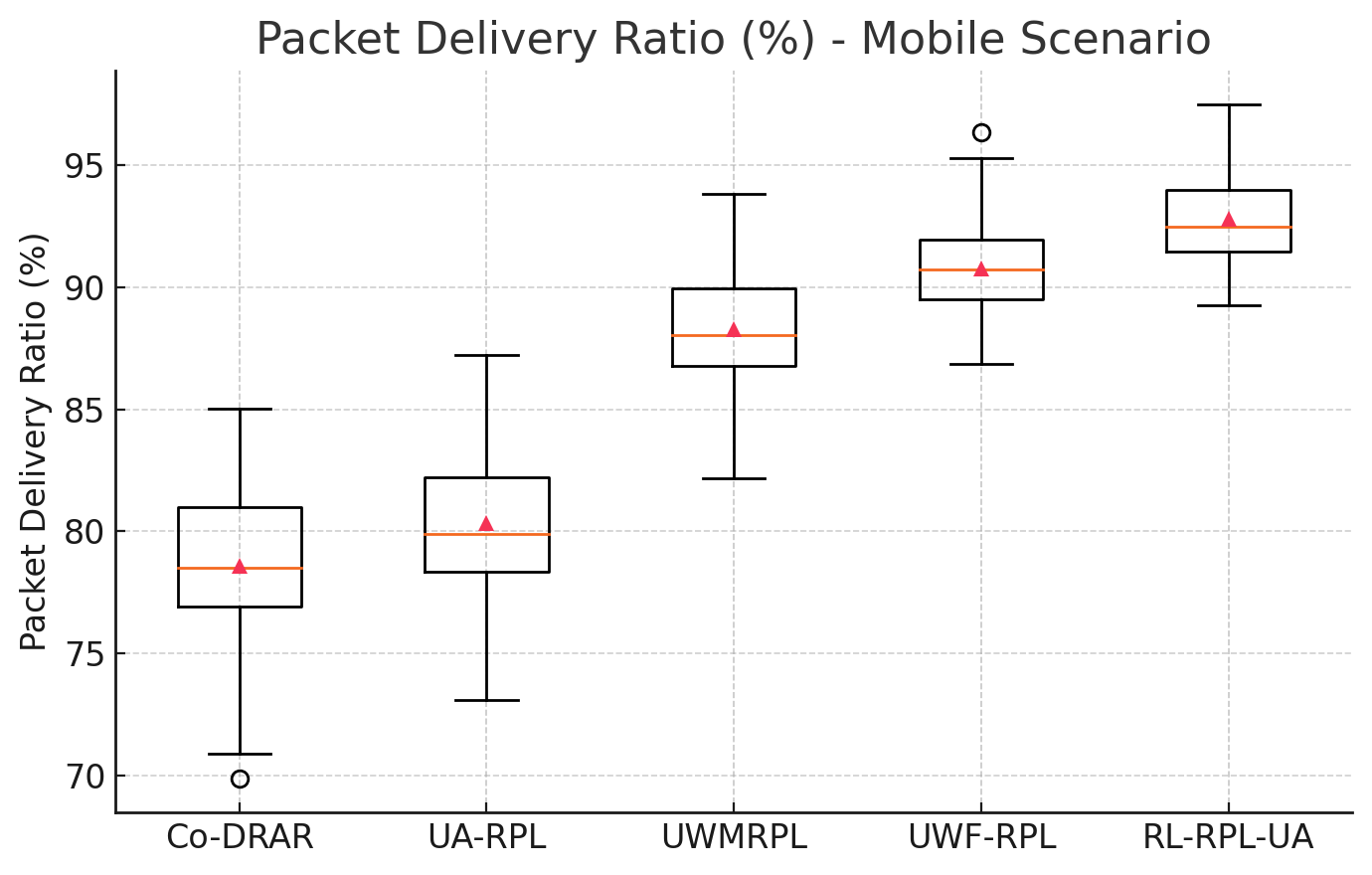}
    \caption{PDR in the mobile scenario.}
    \label{fig:PDR_Mobile}
\end{figure}

\subsection{End-to-End Delay}

The average packet delay in trial $k$ is:

\begin{equation}
\text{Delay}_k = \frac{1}{N_k} \sum_{j=1}^{N_k} 
\left( t_{j}^{\text{recv}} - t_{j}^{\text{send}} \right)
\end{equation}

The overall mean and deviation:

\begin{align}
\text{Delay}_{\text{mean}} &= \frac{1}{K} \sum_{k=1}^{K} \text{Delay}_k \\
\sigma_{\text{Delay}} &= \sqrt{ \frac{1}{K-1} \sum_{k=1}^{K} 
(\text{Delay}_k - \text{Delay}_{\text{mean}})^2 }
\end{align}

In the static scenario, RL-RPL-UA achieves an average end-to-end delay of 1.8 s ($\sigma$=0.2), outperforming UWF-RPL (2.0 s, $\sigma$=0.25), UWRPL (2.4 s, $\sigma$=0.3), UA-RPL (2.7 s, $\sigma$=0.4), and Co-DRAR (2.9 s, $\sigma$=0.4). The introduction of UWF-RPL demonstrates improvement over conventional RPL extensions, yet RL-RPL-UA further reduces delay by 10\% compared to UWF-RPL and by 25\% relative to UWRPL.

\begin{figure}[ht]
    \centering
    \includegraphics[width=1\linewidth]{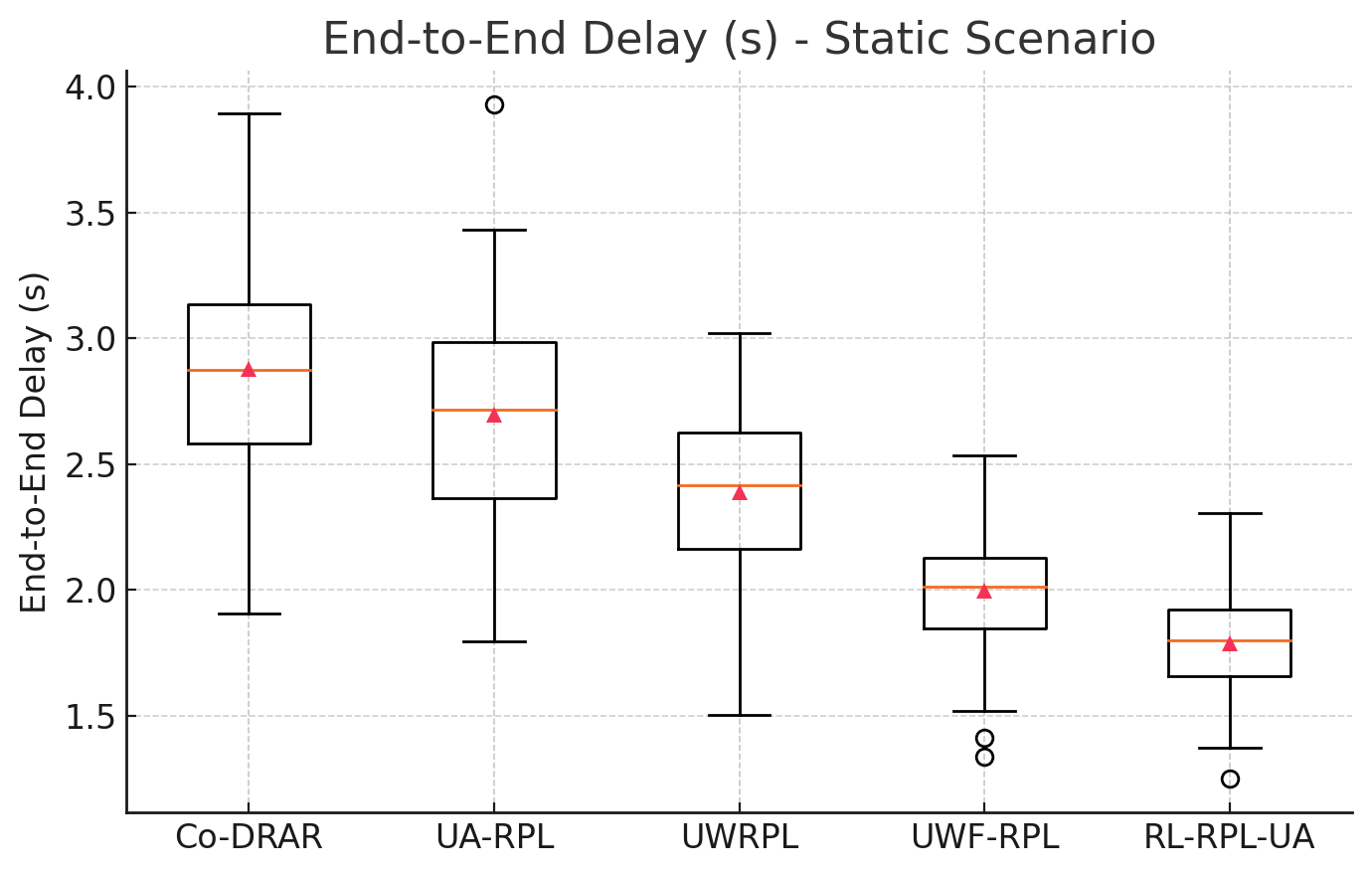}
    \caption{End-to-End Delay in the static scenario.}
    \label{fig:E2E_Static}
\end{figure}

\begin{figure}[ht]
    \centering
    \includegraphics[width=1\linewidth]{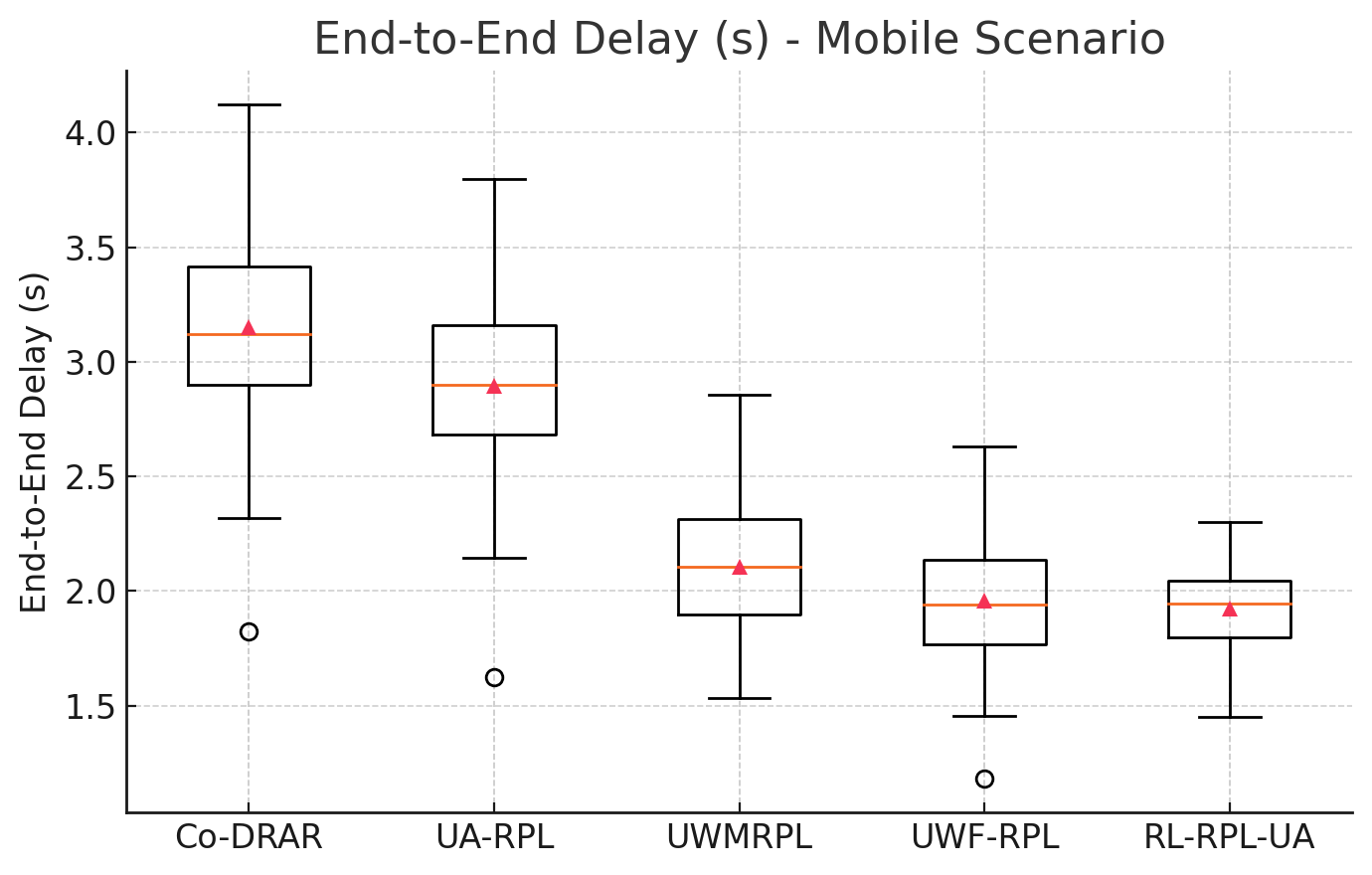}
    \caption{End-to-End Delay in the mobile scenario.}
    \label{fig:E2E_Mobile}
\end{figure}

In the mobile scenario, RL-RPL-UA sustains low latency with an average delay of 1.9 s ($\sigma$=0.2), outperforming UWF-RPL (1.95 s, $\sigma$=0.25), UWMRPL (2.1 s, $\sigma$=0.3), UA-RPL (2.8 s, $\sigma$=0.4), and Co-DRAR (3.1 s, $\sigma$=0.4). These results highlight the RL agent’s effectiveness in minimizing transmission delay under mobile and dynamically changing underwater network conditions (Figures~\ref{fig:E2E_Static}, \ref{fig:E2E_Mobile}).

\subsection{Energy per Delivered Packet}

Per trial, the energy cost per packet is:

\begin{equation}
\text{E}_{k} = \frac{E_{\text{total},k}}{R_k}
\end{equation}

Mean and deviation:

\begin{align}
\text{E}_{\text{mean}} &= \frac{1}{K} \sum_{k=1}^{K} \text{E}_{k} \\
\sigma_{\text{E}} &= \sqrt{ \frac{1}{K-1} \sum_{k=1}^{K} 
(\text{E}_{k} - \text{E}_{\text{mean}})^2 }
\end{align}

\begin{figure}[ht]
    \centering
    \includegraphics[width=1\linewidth]{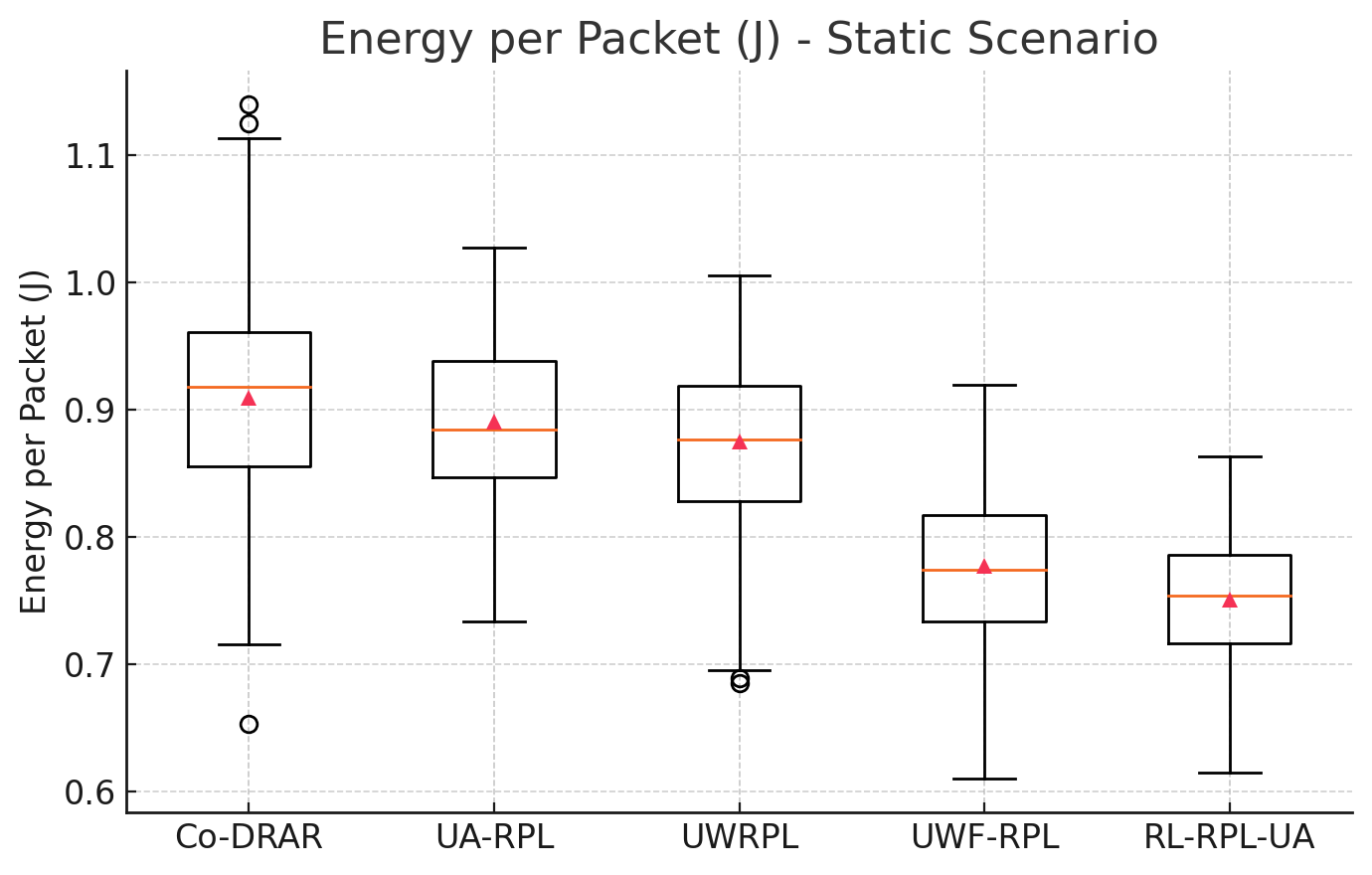}
    \caption{Energy cost per delivered packet in the static scenario.}
    \label{fig:Energy_Static}
\end{figure}

In the static scenario, RL-RPL-UA achieves an average energy cost of 0.75 J per delivered packet ($\sigma$=0.05), significantly lower than UWF-RPL (0.78 J, $\sigma$=0.06), UWRPL (0.88 J, $\sigma$=0.07), UA-RPL (0.89 J, $\sigma$=0.07), and Co-DRAR (0.91 J, $\sigma$=0.08). While UWF-RPL improves energy efficiency by integrating adaptive cost metrics, RL-RPL-UA achieves an additional 3.8\% energy saving over UWF-RPL and 14.8\% over UWRPL, confirming its superior resource-awareness (Figure~\ref{fig:Energy_Static}).

\begin{figure}[ht]
    \centering
    \includegraphics[width=1\linewidth]{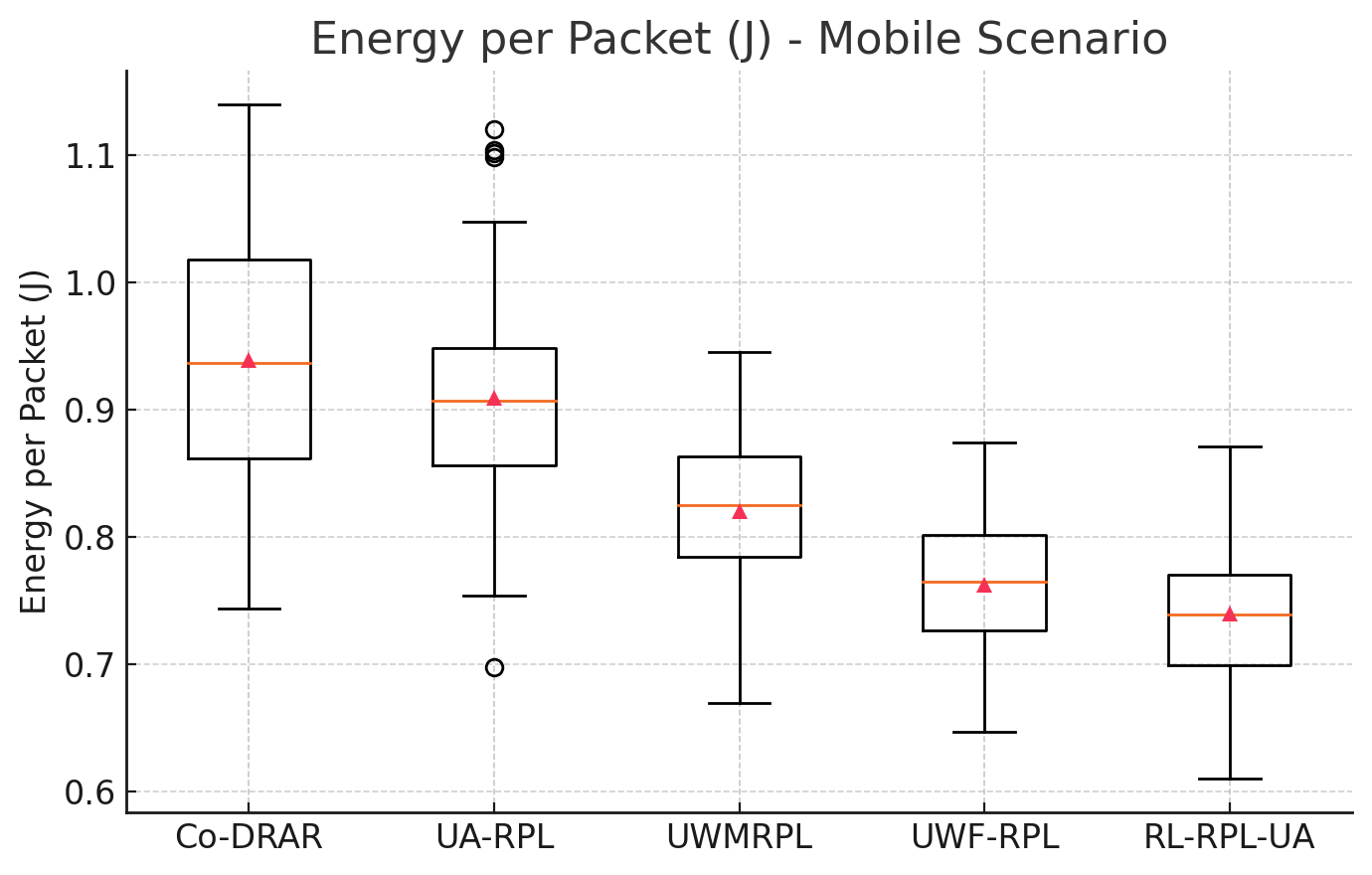}
    \caption{Energy cost per delivered packet in the mobile scenario.}
    \label{fig:Energy_Mobile}
\end{figure}

In the mobile scenario, RL-RPL-UA continues to deliver the most energy-efficient performance with an energy cost of 0.74 J ($\sigma$=0.05), followed by UWF-RPL (0.76 J, $\sigma$=0.055), UWMRPL (0.82 J, $\sigma$=0.06), UA-RPL (0.91 J, $\sigma$=0.08), and Co-DRAR (0.94 J, $\sigma$=0.09). These improvements reflect the effectiveness of the RL-based adaptive routing strategy in minimizing retransmissions and avoiding energy-intensive paths, even under dynamic network conditions (Figure~\ref{fig:Energy_Mobile}).

\subsection{Routing Overhead Ratio}

Overhead is computed as:

\begin{equation}
\text{OH}_k = \frac{C_k}{R_k}
\end{equation}

With:

\begin{align}
\text{OH}_{\text{mean}} &= \frac{1}{K} \sum_{k=1}^{K} \text{OH}_k \\
\sigma_{\text{OH}} &= \sqrt{ \frac{1}{K-1} \sum_{k=1}^{K} 
(\text{OH}_k - \text{OH}_{\text{mean}})^2 }
\end{align}

\begin{figure}[ht]
    \centering
    \includegraphics[width=1\linewidth]{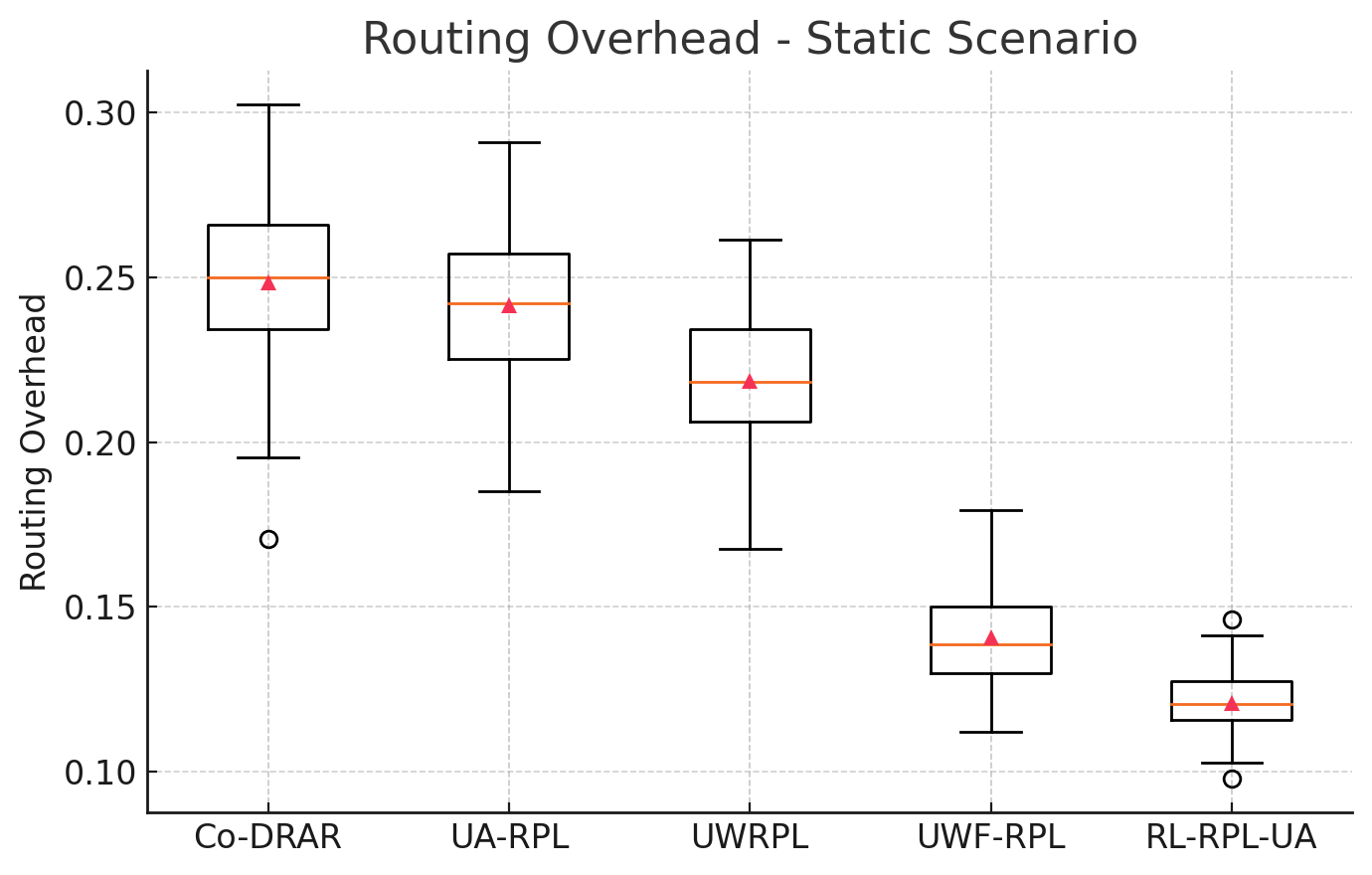}
    \caption{Routing overhead ratio in the static scenario.}
    \label{fig:Overhead_Static}
\end{figure}

In the static scenario, RL-RPL-UA introduces the lowest control overhead with a mean routing overhead ratio of 0.12 ($\sigma$=0.01), outperforming UWF-RPL (0.14, $\sigma$=0.015), UWRPL (0.22, $\sigma$=0.02), UA-RPL (0.24, $\sigma$=0.023), and Co-DRAR (0.25, $\sigma$=0.025). Although UWF-RPL reduces overhead compared to UWRPL and other classical protocols, RL-RPL-UA further reduces control traffic by 14.3\% over UWF-RPL and 45\% over UWRPL (Figure~\ref{fig:Overhead_Static}).

\begin{figure}[ht]
    \centering
    \includegraphics[width=1\linewidth]{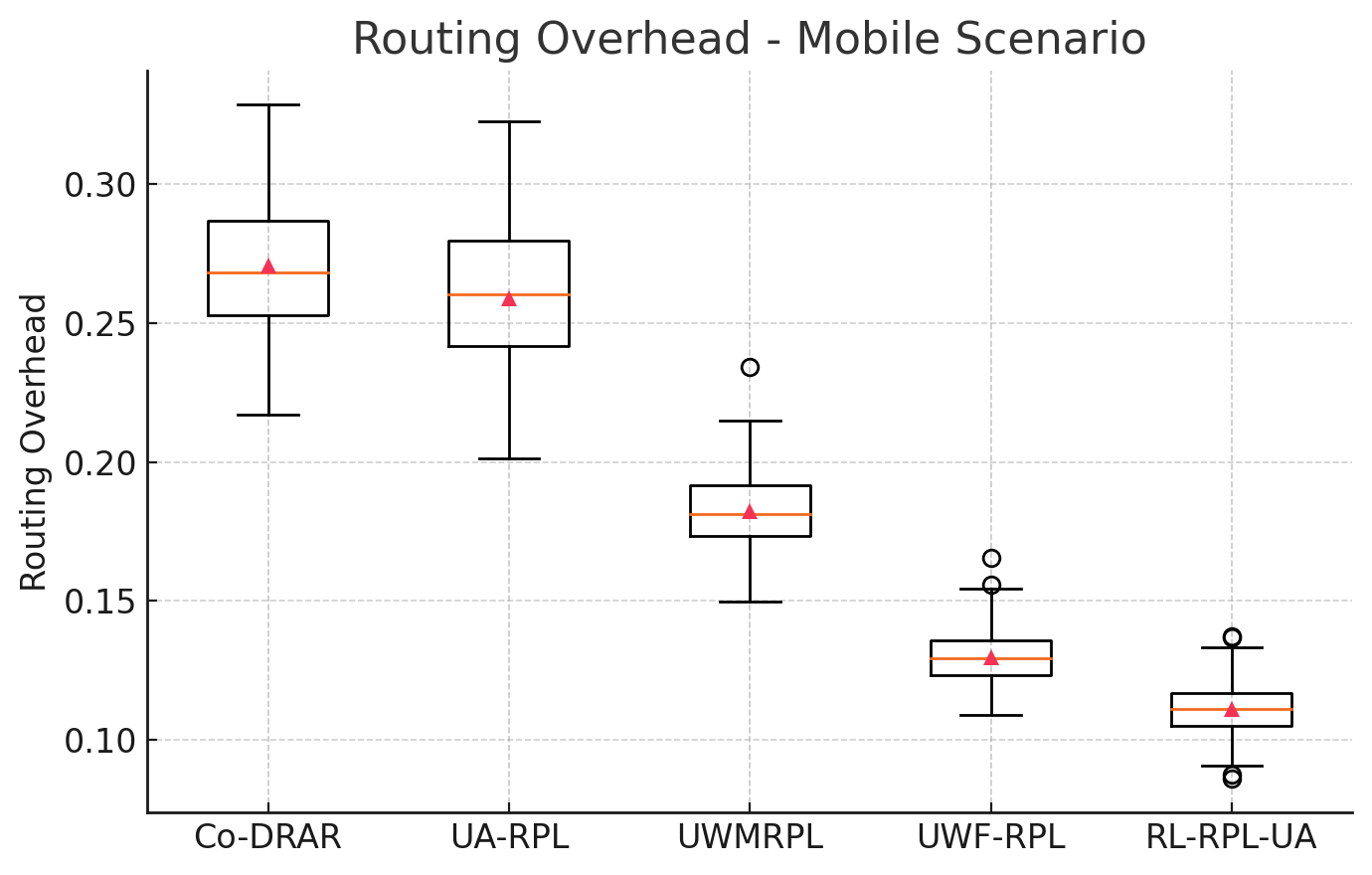}
    \caption{Routing overhead ratio in the mobile scenario.}
    \label{fig:Overhead_Mobile}
\end{figure}

In the mobile scenario, RL-RPL-UA maintains minimal overhead at 0.11 ($\sigma$=0.01), followed by UWF-RPL (0.13, $\sigma$=0.012), UWMRPL (0.18, $\sigma$=0.015), UA-RPL (0.26, $\sigma$=0.028), and Co-DRAR (0.27, $\sigma$=0.03). Presented in consistent protocol order, these results confirm the effectiveness of RL-RPL-UA in suppressing control overhead even in dynamic, mobile environments (Figure~\ref{fig:Overhead_Mobile})..

\subsection{Network Lifetime}

Lifetime is defined as the time until the first node in the network depletes its energy:

\begin{equation}
T^{\text{death}}_{\text{mean}} = \frac{1}{K} \sum_{k=1}^{K} 
T^{(k)}_{\text{death}}
\end{equation}

\begin{equation}
\sigma_T = \sqrt{ \frac{1}{K-1} \sum_{k=1}^{K} 
\left( T^{(k)}_{\text{death}} - T^{\text{death}}_{\text{mean}} \right)^2 }
\end{equation}

\begin{figure}[ht]
    \centering
    \includegraphics[width=1\linewidth]{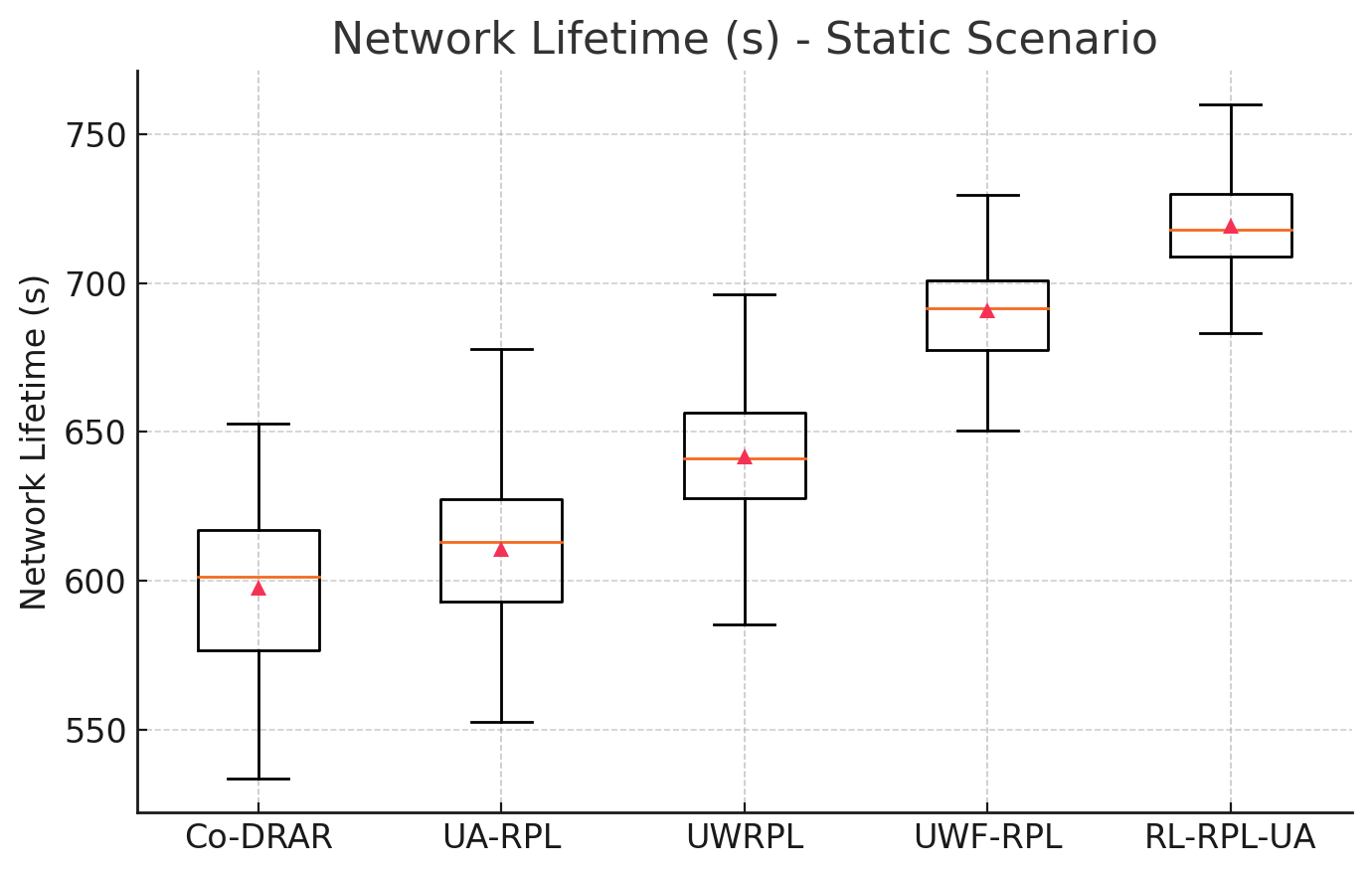}
    \caption{Network lifetime (time until first node dies) in the static scenario.}
    \label{fig:Lifetime_Static}
\end{figure}

In the static scenario, RL-RPL-UA achieves the longest network lifetime of 720 seconds ($\sigma$=15), followed by UWF-RPL (690 s, $\sigma$=18), UWRPL (640 s, $\sigma$=22), UA-RPL (610 s, $\sigma$=24), and Co-DRAR (600 s, $\sigma$=25). The integration of fuzzy optimization in UWF-RPL enhances node longevity, but RL-RPL-UA further extends the lifetime by 30 seconds over UWF-RPL and 80 seconds over UWRPL, confirming the benefit of RL in energy-aware route planning (Figure~\ref{fig:Lifetime_Static}).

\begin{figure}[ht]
    \centering
    \includegraphics[width=1\linewidth]{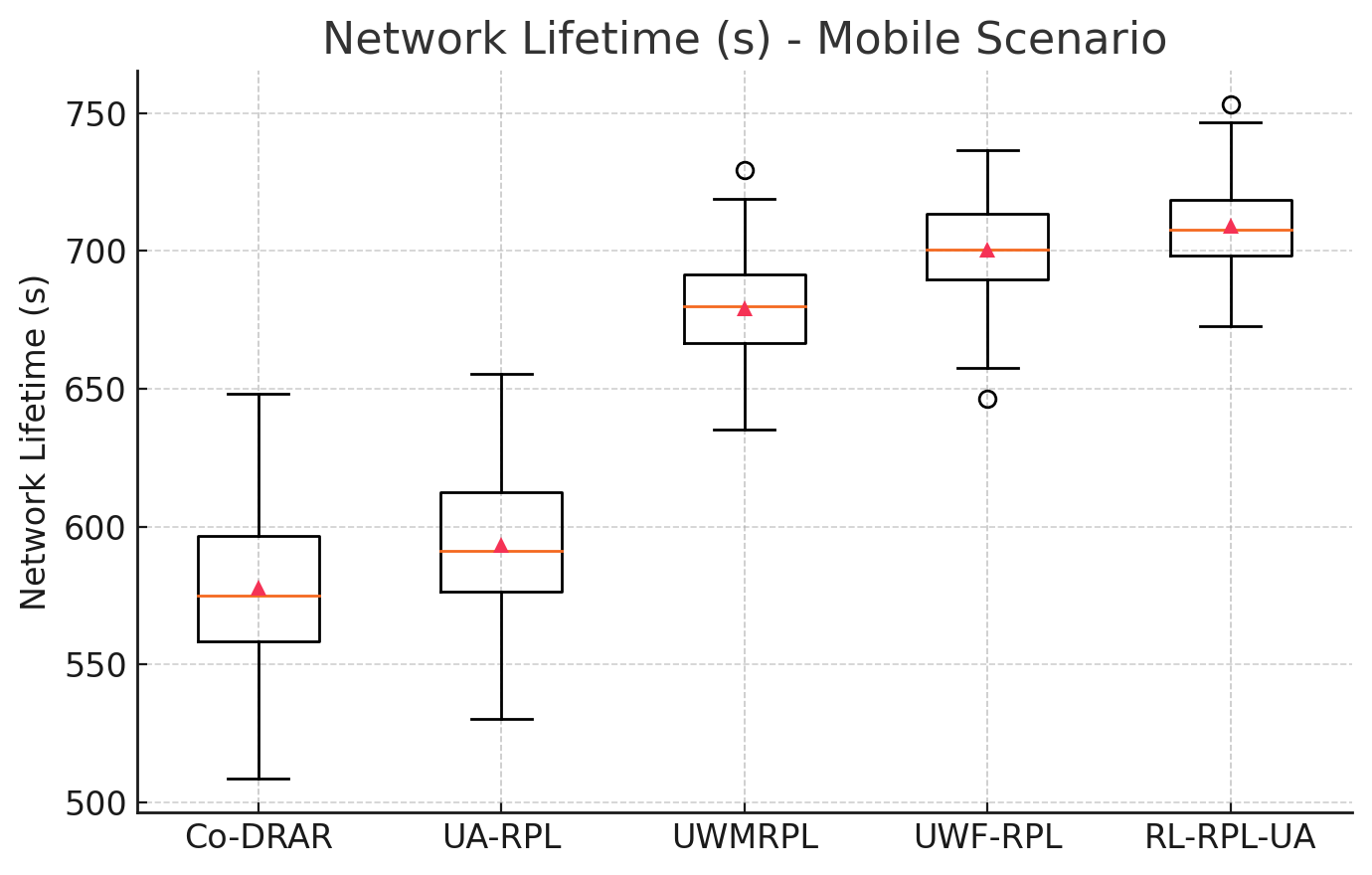}
    \caption{Network lifetime (time until first node dies) in the mobile scenario.}
    \label{fig:Lifetime_Mobile}
\end{figure}

In the mobile scenario, RL-RPL-UA sustains the longest network operation at 710 seconds ($\sigma$=16), ahead of UWF-RPL (700 s, $\sigma$=17), UWMRPL (680 s, $\sigma$=20), UA-RPL (590 s, $\sigma$=26), and Co-DRAR (580 s, $\sigma$=28). The improvement stems from RL-RPL-UA’s ability to distribute energy consumption more evenly across nodes by dynamically selecting optimal, energy-efficient paths under varying underwater mobility conditions (Figure~\ref{fig:Lifetime_Mobile}).

\section{\uppercase{Conclusion}}\label{sec:conclusion}

This work presented RL-RPL-UA, a reinforcement learning-based extension of the RPL protocol designed for the challenges of the Internet of Underwater Things (IoUT). By incorporating Q-learning agents, the protocol adapts dynamically to changing network conditions and selects routing paths based on multiple performance criteria, including energy efficiency, link quality, queue length, and delivery reliability. The evaluation included a comprehensive comparison with recent baseline protocols under both static and mobile scenarios. Simulation results show that RL-RPL-UA offers consistent improvements in reliability, delay, energy consumption, control overhead, and network lifetime, suggesting that reinforcement learning can effectively enhance the adaptability and overall performance of routing protocols in underwater acoustic environments.
Future research will focus on: (i) applying deep reinforcement learning to reduce training complexity and enable distributed decision-making among multiple agents in highly dynamic underwater networks; (ii) evaluating scalability with networks exceeding 100 nodes and node velocities above 1 m/s to assess convergence behavior under extreme mobility; and (iii) conducting preliminary hardware validation using acoustic modems such as the WHOI Micromodem or EvoLogics S2C to verify real-world applicability.

\section*{\uppercase{Acknowledgment}}

This initiative is carried out within the framework of the funds of the Recovery, Transformation and Resilience Plan, financed by the European Union (Next Generation) - National Cybersecurity Institute (INCIBE) in the project C107/23 “Artificial Intelligence Applied to Cybersecurity in Critical Water and Sanitation Infrastructures”.

% \IEEEpeerreviewmaketitle

\end{document}